\documentclass{article}
\usepackage{amsmath}
\usepackage{amsfonts}
\usepackage{amssymb}

\font\small=cmr8
\input{tcilatex}

\begin{document}

\title{Remarks on Robin's and Nicolas Inequalities}
\author{Boris A. Kupershmidt \\
The University of Tennessee Space Institute,\\
Tullahoma, TN 37388\\
bkupershmidt@utsi.edu}
\maketitle

\begin{abstract}
Nicolas Conjecture is disproved. The Robin Conjecture follows.
\end{abstract}

\begin{flushright}
{\small{To V. I. Arnold, the greatest mathematician of all times.}}\\ 
\end{flushright}

\begin{flushright}
{\small{I believe this to be false.  There is no evidence whatever for it (unless one 
counts that it is always nice when any function has only real roots). One should not
believe things for which there is no evidence.  In the spirit of this anthology I should
also record my feelings that there is no imaginable reason why it should be true.\\ 
J. E. Littlewood on Riemnann's Conjecture}} \\ 
\end{flushright}

\begin{flushright}
{\small{... We are all in our own eyes a failure:  after all, \\
we haven't proved Fermat's Last Theorem, \\ 
nor Riemann's Conjecture. \\
Mary L. Cartwright}} \\
\end{flushright}

\section{Introduction}

The Nicolas Conjecture [Nic 1983] states that 
\begin{equation}
\frac{{\mathcal{N}}_{k}}{\varphi ({\mathcal{N}}_{k})}>e^{\gamma }\log \log {%
\mathcal{N}}_{k},\ \ k\geq 1,\   \tag{0}
\end{equation}%
where: 
\begin{equation}
{\mathcal{N}}_{k}=\prod_{i=1}^{k}p_{i},\   \notag
\end{equation}%
$p_{i}$ is the prime number $\#i$, $\varphi $ is the Euler phi-function, and 
$\gamma =0.57...$ is the Euler constant. For more details, see the beautiful
paper [CLM 2006], where it was proven that 
\begin{equation}
\prod_{i=1}^{k}(p_{i}+1)/p_{i}<e^{\gamma }\log \log {\mathcal{N}}_{k}^{2},\
k>4,  \notag
\end{equation}%
in contrast to (1): 
\begin{equation}
\prod_{i=1}^{k}p_{i}/(p_{i}-1)>e^{\gamma }\log \log {\mathcal{N}}_{k},\
k\geq 1.  \notag
\end{equation}

Crucially, Nicolas proved that if his Conjecture is \textit{not} true then
the inequality (1) is both true and untrue infinitely often. Thus, it's
enough to establish it for $k > > 1$, i.e. for 
$k$ large enough.

The Robin's inequality, equivalent to Nicolas one, is: 
\begin{equation}
\frac{\sigma (n)}{n}<e^{\gamma }\log \log n,\ \ n\geq 5041,  \tag{1}
\end{equation}%
where $\sigma (n)=\sigma _{1}(n)=\sum_{d|n}d$ is the sum of divisors of $n$.
[Rob 1984].

As is clear from the table of exceptions $\leq $ 5040 in [CLM 2006], (1) can
be replaced by: 
\begin{equation}
\frac{\sigma (n)}{n}<e^{\gamma }\log \log n,\ \ \omega (n)>4,  \tag{2}
\end{equation}%
where $\omega (n)$ is the number of distinct prime divisors of $n$.

CLMS also mention that it's enough to consider only those $n$ which are in
Hardy-Ramanujan form: if $\omega (n) = m$, 
\begin{equation}
n = p_1^{e_{1}} . . . p_m^{e_{m}},  \tag{3}
\end{equation}
then 
\begin{equation}
e_1 \geq e_2 \geq . . . \geq e_m. \   \notag
\end{equation}

Still, the inequality (1) has arbitrary many parameters: the $e_i$'s. This
is not conducive to a proof. The Nicolas conjecture offers better chances.

\section{The Method}

Thus, in handling the Nicolas inequality, we need 
\begin{equation}
\log LHS\overset{?}{>}\log RHS,\   \notag
\end{equation}%
where 
\begin{gather}
\log LHS=\sum_{i=1}^{m}\log \bigg(1+\frac{1}{p_{i}-1}\bigg),  \tag{4} \\
\log RHS=\gamma +\log \log \theta (p_{m}).\   \tag{5}
\end{gather}

Now, 
\begin{gather}
\theta (p_{m})\leq p_{m}\bigg(1+\frac{\eta _{s}}{\log ^{s}p_{m}}\bigg),\
s=1,2,3\ \Rightarrow   \notag \\
\log \theta (p_{m})\leq \log p_{m}+\frac{\eta _{s}}{\log ^{s}p_{m}}=\log
p_{m}\bigg(1+\frac{\eta _{s}}{\log ^{s+1}p_{m}}\bigg)\Rightarrow   \notag \\
\log \log \theta (p_{m})\leq \log \log p_{m}+\frac{\eta _{s}}{\log
^{s+1}p_{m}},  \tag{6}
\end{gather}%
and we are going to look, beyond first 2 terms, $\gamma +\log \log p_{m}$,
at the series in $\frac{1}{\log p_{m}}$. Thus, 
\begin{equation}
\log RHS=\log \log p_{m}+\gamma +\frac{\eta _{3}}{\log ^{4}p_{m}}+... 
\tag{7}
\end{equation}

For the LHS, we have: 
\begin{gather}
\sum_{i=1}^{n}\log \bigg(1+\frac{1}{p_{i}-1}\bigg)=\sum_{i=1}^{m}\bigg\{%
\bigg[\log \bigg(1+\frac{1}{p_{i}-1}\bigg)-\frac{1}{p_{i}}\bigg]+\frac{1}{%
p_{i}}\bigg\}=  \notag \\
=\sum_{i=1}^{m}\bigg[\log \bigg(1+\frac{1}{p_{i}-1}\bigg)-\frac{1}{p_{i}}%
\bigg]+\sum_{i=1}^{m}\frac{1}{p_{i}}=\sum_{i-1}^{m}\frac{1}{p_{i}}%
+\sum_{i=1}^{\infty }\log \bigg(1+\frac{1}{p_{i}-1}\bigg)-\frac{1}{p_{i}}%
\bigg]-  \notag \\
-\sum_{m+1}^{\infty }\bigg[\log \bigg(1+\frac{1}{p_{i}-1}\bigg)-\frac{1}{%
p_{i}}\bigg]\bigg(\bigg[\mathrm{Fin}\ 2003\bigg],\mathrm{p}.95\bigg)=  \notag
\\
=\sum_{i=1}^{m}\frac{1}{p_{i}}+(\gamma -M)-\sum_{m+1}^{\infty }\log \bigg(1+%
\frac{1}{p_{i}-1}\bigg)-\frac{1}{p_{i}}\bigg],  \tag{8}
\end{gather}%
where ([Fin 2003], p. 95) 
\begin{gather}
\sum_{i=1}^{m}\frac{1}{p_{i}}=\log \log p_{m}+M+0(1),  \tag{9} \\
M=0.261497...\   \tag{10}
\end{gather}

Let's dispose now of $\sum_{m+1}^{\infty }$ - term. We have: 
\begin{equation}
\log \bigg(1+\frac{1}{p-1}\bigg)-\frac{1}{p}\sim \frac{1}{p-1}-\frac{1}{p}%
\sim \frac{1}{i^{2}\log ^{2}i}\   \notag
\end{equation}%
and 
\begin{equation}
\int_{m+1}^{\infty }\frac{dx}{x^{2}\log ^{2}x}\sim -\frac{1}{x\log ^{2}x}%
\bigg|_{m+1}^{\infty }\sim \frac{1}{m\log ^{2}m},\   \notag
\end{equation}%
and we are going to count only $m$-free terms in comparing the log LHS with
the log RHS.

Thus, the first two leading terms: $\log \log p_{m}$ and $\gamma $ - being
equal on both sides, we are comparing 
\begin{equation}
\sum_{i=1}^{m}\frac{1}{p_{i}}-\log \log p_{m}-M\ \mathrm{and}\ 0,  \tag{11}
\end{equation}%
at least modulo 
\begin{equation}
\frac{1}{\log ^{3}m}.  \tag{12}
\end{equation}

We need the first nonzero term in the log LHS, in (11), being $> 0$ (or $< 0$%
, as the case may be).

\section{The Proof}

We calculate modulo $1/m^{2}$ and modulo $1/m\log ^{4}m$. Our coefficients
are polynomial functions in $w=\log \log m$. \newline

\noindent \textbf{Lemma 13.} 
\begin{equation}
\log (m+1)=\log m+\frac{1}{m}.  \tag{14}
\end{equation}%
\noindent \textbf{Proof.} 
\begin{equation}
\log (m+1)=\log [m\bigg(1+\frac{1}{m}\bigg)]=\log m+\log \bigg(1+\frac{1}{m}%
\bigg)\equiv \log m+\frac{1}{m}\ \ \blacksquare   \notag
\end{equation}%
\noindent \textbf{Lemma 15.} 
\begin{equation}
\log \log (m+1)=\log \log m+\frac{1}{m\log m}.  \tag{16}
\end{equation}%
\noindent \textbf{Proof.} 
\begin{gather}
\log \log (m+1)=\log \bigg(\log m+\frac{1}{m}\bigg)\equiv \log \bigg[\log m%
\bigg(1+\frac{1}{m\log m}\bigg)\bigg]=  \notag \\
=\log \log m+\frac{1}{m\log m}.\ \ \ \blacksquare 
\end{gather}

Let $C=C(w),w=\log \log m$. Call it $C_{m}$. \newline

\noindent \textbf{Lemma 17.} 
\begin{equation}
\frac{C_{m}}{\log ^{k}m}-\frac{C_{m+1}}{\log ^{k}(m+1)}=\frac{kC-C^{\prime }%
}{m\log ^{k+1}m}.  \tag{18}
\end{equation}%
\noindent \textbf{Proof.} We have: 
\begin{gather}
LHS=\frac{C}{\log ^{k}m}-\frac{C+C^{\prime }\frac{1}{m\log m}}{\bigg(\log m+%
\frac{1}{m}\bigg)^{k}}=\frac{1}{\log ^{k}m\bigg(\log ^{k}m+\frac{k}{m}\log
^{k-1}m\bigg)}\ \mathrm{times}:  \notag \\
C\bigg(\log ^{k}m+\frac{k}{m}\log ^{k-1}m\bigg)-\bigg(C+C^{\prime }\frac{1}{%
m\log m}\bigg)\log ^{k}m=  \notag \\
=\frac{\log ^{k-1}m}{m}(kC-C^{\prime }).\   \notag
\end{gather}%
\noindent Altogether, 
\begin{equation}
LHS=\frac{1}{m\log ^{k+1}m}\ (kC-C^{\prime }).\ \ \blacksquare \   \notag
\end{equation}%
\noindent \textbf{Lemma 19.} 
\begin{equation}
\frac{1}{p_{m+1}}=\frac{1}{p_{m}}.  \tag{20}
\end{equation}%
\noindent \textbf{Proof.} Set 
\begin{equation}
p_{m}=mf(m),  \tag{21}
\end{equation}%
where 
\begin{gather}
f(m)=\log m+(\log \log m-1)+\frac{\log \log m-2}{\log n}+...=  \notag \\
=\log m+\sum_{i\geq 0}\frac{P_{i}}{\log ^{i}m}.  \tag{22}
\end{gather}%
Denote $P(m+1)=\tilde{P}$. Then 
\begin{equation}
\frac{1}{p_{m}}-\frac{1}{p_{m+1}}=\frac{1}{p_{m}p_{m+1}}(p_{m+1}-p_{m})\sim 
\frac{1}{m^{2}\log ^{2}m}(p_{m+1}-p_{m}),  \tag{23}
\end{equation}%
and 
\begin{gather}
p_{m+1}-p_{m}=(m+1)f(m+1)-mf(m)=  \notag \\
=m[f(m+1)-f(m)]+f(m+1)\equiv m[f(m+1)-f(m)],\   \notag
\end{gather}%
and 
\begin{equation}
f(m+1)-f(m)=\log (m+1)-\log m+\sum_{i\geq 1}\bigg(\frac{\tilde{P}_{i}}{\log
^{i}(m+1)}-\frac{P_{i}}{\log ^{i}m}\bigg).\   \notag
\end{equation}%
\noindent But 
\begin{equation}
\log (m+1)-\log m=\frac{1}{m},\   \notag
\end{equation}%
and by (18), 
\begin{equation}
\frac{\tilde{P}_{i}}{\log ^{i}(m+1)}-\frac{P_{i}}{\log ^{i}m}=\frac{%
P_{i}^{\prime }-iP_{i}}{m\log ^{i+1}m}.\   \notag
\end{equation}%
Together with (23), this proves (20). \ \ $\blacksquare $

We are interested in $C=C_{1},D=C_{2}$, such that 
\begin{equation}
\sum_{i=1}^{m}\frac{1}{p_{i}}=\log \log p_{m}+M+\frac{C}{\log m}+\frac{D}{%
\log ^{2}m}+...\   \notag
\end{equation}%
We use induction on $m$ to find $C$ and $D$. We have: 
\begin{gather}
\sum_{i=1}^{m}\frac{1}{p_{i}}+\frac{1}{p_{m+1}}=\frac{1}{p_{m+1}}+\log \log
p_{m}+M+\frac{C}{\log m}+\frac{D}{\log ^{2}m}=  \notag \\
=\log \log p_{m+1}+M+\frac{\tilde{C}}{\log (m+1)}+\frac{\tilde{D}}{\log
^{2}(m+1)},  \tag{24}
\end{gather}%
or 
\begin{equation}
\bigg(\frac{C}{\log m}-\frac{\tilde{C}}{\log (m+1)}\bigg)+\bigg(\frac{D}{%
\log ^{2}m}-\frac{\tilde{D}}{\log ^{2}(m+1)}\bigg)=(\log \log p_{m+1}-\log
\log p_{m})-\frac{1}{p_{m}}.  \tag{25}
\end{equation}%
By (18), this is: 
\begin{equation}
\frac{C-C^{\prime }}{m\log ^{2}m}+\frac{2D-D^{\prime }}{m\log ^{3}m}=(\log
\log p_{m+1}-\log \log p_{m})-\frac{1}{p_{m}}.  \tag{26}
\end{equation}

Let's start with the $1/p_{m}-$ term. We have: 
\begin{gather}
-\frac{1}{p_{m}}=-\frac{1}{m\log m}\bigg(\frac{1}{1+\frac{P_{0}}{\log m}+%
\frac{P_{1}}{\log ^{2}m}}\bigg)=  \notag \\
=-\frac{1}{m\log m}\bigg(1-\frac{P_{0}}{\log m}-\frac{P_{1}}{\log ^{2}m}+%
\frac{P_{0}^{2}}{\log ^{2}m}\bigg).\   \tag{27}
\end{gather}%
\noindent \textbf{Lemma 28.} 
\begin{equation}
\log p_{m+1}-\log p_{m}\sim \frac{1}{m}\bigg(1+\frac{1}{\log m}\bigg)-\frac{%
P_{0}^{\prime }-P_{0}}{m\log ^{2}m}.  \tag{29}
\end{equation}%
\textbf{Proof.} We have: 
\begin{gather}
LHS=\log \frac{p_{m+1}}{p_{m}}=\log \bigg[\frac{m+1}{m}\frac{f(m+1)}{f(m)}%
\bigg]=\frac{1}{m}+\log \frac{f(m+1)}{f(m)}=  \notag \\
=\frac{1}{m}+\log \bigg\{\frac{\log (m+1)}{\log m}\bigg[\frac{1+\frac{\tilde{%
P}_{0}}{\log (m+1)}+\frac{\tilde{P}_{1}}{\log ^{2}(m+1)}+...}{1+\frac{P_{0}}{%
\log m}+\frac{\tilde{P}_{1}}{\log ^{2}m}+...}\bigg]\bigg\}=  \notag \\
=\frac{1}{m}+\frac{1}{m\log m}+\log \bigg[\frac{\ .\ .\ .\ }{\ .\ .\ .\ }%
\bigg]=  \notag \\
=\frac{1}{m}\bigg(1+\frac{1}{\log m}\bigg)+\log \bigg[1+\frac{\tilde{P}_{0}}{%
\log (m+1)}+\frac{\tilde{P}_{1}}{\log ^{2}(m+1)}+...\bigg]-\log \bigg[1+%
\frac{P_{0}}{\log m}+\frac{P_{1}}{\log ^{2}m}+...\bigg]=  \notag \\
=\frac{1}{m}\bigg(1+\frac{1}{\log m}\bigg)+\frac{\tilde{P}_{0}}{\log (m+1)}-%
\frac{P_{0}}{\log m}\ \ [\mathrm{by}\ (18)]=  \notag \\
=\frac{1}{m}\bigg(1+\frac{1}{\log m}\bigg)-\frac{P_{0}^{\prime }-P}{m\log
^{2}m}.\ \ \blacksquare \   \notag
\end{gather}

\noindent \textbf{Lemma 30.} 
\begin{equation}
\frac{1}{\log p_{m}}=\frac{1}{\log m}\bigg(1-\frac{\log \log m}{\log m}\bigg)%
+O\bigg(\frac{1}{\log ^{3}m}\bigg).  \tag{31}
\end{equation}%
\noindent \textbf{Proof.} We have: 
\begin{gather}
\frac{1}{\log p_{m}}=\frac{1}{\log m+\log f(m)}=\frac{1}{\log m}\frac{1}{[1+%
\frac{1}{\log m}\log f(m)]}=  \notag \\
=\frac{1}{\log m}\bigg(1-\frac{\log \log m}{\log m}\bigg)\ \ \ \blacksquare
\   \notag
\end{gather}%
\noindent \textbf{Lemma 32.} 
\begin{equation}
\log \log p_{m+1}-\log \log p_{m}=\frac{1}{m\log m}\bigg(1+\frac{1}{\log m}%
\bigg)+\frac{P_{0}^{\prime }-P_{0}}{m\log ^{3}m}-\frac{\log \log m}{m\log
^{2}m}\bigg(1+\frac{1}{\log m}\bigg).  \tag{33}
\end{equation}%
\noindent \textbf{Proof.} We have: 
\begin{gather}
LHS=\log \frac{\log p_{m+1}}{\log p_{m}}=\log \bigg(1+\frac{\log
p_{m+1}-\log p_{m}}{\log p_{m}}\bigg)=  \notag \\
=\frac{\log p_{m+1}-\log p_{m}}{\log p_{m}}[\mathrm{by}\ (29),(31)]=  \notag
\\
=\frac{1}{\log m}\bigg(1-\frac{\log \log m}{\log m}\bigg)\bigg\{\frac{1}{m}%
\bigg(1+\frac{1}{\log m}\bigg)-\frac{P_{0}^{\prime }-P_{0}}{m\log ^{2}m}%
\bigg\}=  \notag \\
=\frac{1}{m\log m}\bigg(1+\frac{1}{\log m}\bigg)-\frac{P_{0}^{\prime }-P_{0}%
}{m\log ^{3}m}-\frac{\log \log m}{m\log ^{2}m}\bigg(1+\frac{1}{\log m}\bigg)%
.\ \ \blacksquare \   \notag
\end{gather}

Collecting all Lemmas, we rewrite (25) as 
\begin{gather}
\frac{C-C^{\prime }}{m\log ^{2}m}+\frac{2D-D^{\prime }}{m\log ^{3}m}=  \notag
\\
=\frac{1}{m\log m}\bigg(1-\frac{P_{0}}{\log m}+\frac{P_{0}^{2}-P_{1}}{\log
^{2}m}\bigg)+  \notag \\
+\frac{1}{m\log m}\bigg(1+\frac{1}{\log m}\bigg)-\frac{P_{0}^{\prime }-P_{0}%
}{m\log ^{3}m}-\frac{\log \log m}{m\log ^{2}m}\bigg(1+\frac{1}{\log m}\bigg).
\tag{34}
\end{gather}%
Thus, 
\begin{gather}
\frac{1}{m\log m}-\mathrm{coefficient\ is\ 0};  \notag \\
\frac{1}{m\log ^{2}m}-\mathrm{coefficient\ is:}  \notag \\
P_{0}+1-\log \log m=0\ \mathrm{because}\ P_{0}=\log \log m-1;  \notag \\
\frac{1}{\log ^{3}m}-\mathrm{coefficient\ is:}  \notag \\
-P_{0}^{2}+P_{1}-P_{0}^{\prime }+P_{0}-\log \log m<0\ \mathrm{for\ }m>>0.\ 
\mathrm{Thus},  \notag \\
LHS<RHS.  \tag{35}
\end{gather}%
The Nicolas Conjecture is thus disproved, but the strengthened Robin
Inequality, is established: 
\begin{equation}
\prod_{i=1}^{m}\frac{p_{i}}{\varphi (p_{i})}<e^{\gamma }\log \log {\mathcal{N%
}}_{m}.\   \notag
\end{equation}%
Notice that Nicolas Inequality is disproved in a way that contradicts the
Nicolas result, that if his inequality is 
wrong, then it's wrong \textit{and \ right} infinitely often. Note that the
equation 
\begin{equation}
kF-F^{\prime }=p(w),\   \notag
\end{equation}%
where $p$ is a polynomial, has trivially a unique polynomial solution in $%
w=\log \log m$.

\section{Acknowledgment 1}

Although not apparent, the paper is a result of hundreds of computer
experiments, and it has an invisible hero: the PARI-GP program, a free to
the mathematical community wonderful tool 
for computer experiments. My deep gratitude to the developers of PARI.

Nicholas Lynn was instrumental in discovering widespread order among the
primes, by running many numerical experiments for me on Mathematica. Later
Tony Saad had replaced him. Joshua 
Kupershmidt set me up with PARI, and helped both with its quirks and the
actual programming.

I thank Remi Engels for technical assistance with Robin's article, and Keith
Briggs for sending me his inspiring paper.

Last but not least, over the years of struggles with the problem, with me
often concluding that it's unsolvable, K. C. Reddy held unshakably to the
firm believe that I'll conquer it. Why he 
was so certain, is a mystery to me.

\section{Acknowledgment 2}

It would be churlish of me not to mention the wonderfully well written and
remarkably exciting and stimulating paper [CLMS 2006].

\section{Acknowledgment 3}

I'm grateful to Pieter Moore for the careful reading of the first version of
this paper and constructing a counterexample.

I'm also deeply indebted to Patrick Sol${\acute{\mathrm{e}}}$ for his
encouragement after reading the fatally flawed 1$^{st}$ version of the
Manuscript.

\end{document}